\documentclass[12pt]{iopart}

\usepackage{amssymb}
\usepackage{epsfig}
\usepackage{graphicx}
\usepackage[colorlinks=true,urlcolor=blue,linkcolor=blue]{hyperref}

\begin{document}

\title{A positive cosmological constant as a geometrical artifact}
\author{Pantelis S. Apostolopoulos$^{1,a}$ and Christos Tsipogiannis$^{1,b}$}

\begin{abstract}
We revisit the classical mechanism to produce $d-$dimensional spacetimes via
Kaluza-Klein compactification. We made the assumption that the $\left(
d+1\right) -$dimensional bulk geometry $\mathcal{M}$ admits a Homothetic
Vector Field (HVF)\ $\mathbf{H}$ relaxing the existence of a zero i.e. $%
\mathbf{H}\neq 0$ holds on $\mathcal{M}$ and the homothetic bivector $F_{ab}(%
\mathbf{H})=0$. Under these circumstances we identify the origin of a
positive cosmological constant in the $d-$dimensional spacetime as the
homothetic factor representing a geometrical artifact of late time state of
the $\left( d+1\right) -$dimensional bulk spacetime.\newline
\newline
Keywords: Homothety; Self-similarity; Kaluza-Klein Compactification;
Conformal Symmetries.
\end{abstract}

\address{$^1$Department of Environment, Ionian University\\
Mathematical Physics and Computational Statistics Research Laboratory\\
Panagoula 29100, Island of Zakynthos, Greece} \ead{$^a$papostol@ionio.gr} %
\ead{$^b$v20tsip@ionio.gr}

\maketitle

\bigskip

\setcounter{equation}{0}

There is a long history regarding the efforts to \textquotedblleft
restore\textquotedblright\ our 4D world from the reduction of higher
dimensional geometries. This program has been initiated by Kaluza \cite%
{Kaluza:1921tu} and Klein \cite{Klein} aiming to introduce a unified theory
of gravity and electromagnetism. The essential idea was that the extra
dimension can be compactified (i.e. is small enough w.r.t the remaining four
spacetime dimensions) and the metric functions are independent from it (the
\textquotedblleft cylinder\textquotedblright\ condition). Later this
condition was relaxed and replaced with a periodic dependence of the fields
where the \textquotedblleft cylinder\textquotedblright\ geometry arises (see
e.g. \cite{Overduin:1997sri}, \cite{PoncedeLeon}, \cite{PaulSWesson}). The
above program has been widely used and new high dimensional theories
appeared. For example, the weakness of gravity at short scales ($\gtrsim 1$%
mm) follows from the existence of compact extra spatial dimensions large
enough compared with the weak scale \cite{Arkani-Hamed:1998jmv}, \cite%
{Antoniadis:1998ig}. At the same time, another theory appeared where,
instead of compactification, the \textquotedblleft return\textquotedblright\
to 4D gravity is achieved if we assume that the spacetime geometry is a
3-brane (hypersurface) embedded in a 5D bulk spacetime and the 3-brane
becomes the boundaries of a finite fifth dimension \cite{Randall:1999ee}, 
\cite{Randall:1999vf}. In the latter case, the bulk spacetime can be either $%
\Lambda -$vacuum (AdS) or with arbitrary matter component where the mirage
term on the 3-brane retains its form, but the black hole mass is replaced by
the covariantly defined integrated mass of the bulk matter \cite%
{Apostolopoulos:2005at}.

In the present work we revisit the above scheme and focus on a geometrical
(symmetry) point of view. Throughout this paper, the following conventions
have been used: the pair ($\mathcal{M},\mathbf{g}$) denotes the $\left(
d+1\right) -$dimensional bulk spacetime manifold endowed with a Lorentzian
metric of signature ($-,+,+,+,...$), bulk $\left( d+1\right) -$dimensional
indices are denoted by capital Latin letters $A,B,...=0,1,2,...,d$ and Greek
letters denote $d-$dimensional indices $\alpha ,\beta
,...=0,1,2,3,...,\left( d-1\right) $.

Self-similar models are characterized by the existence of a global smooth
vector field $\mathbf{H}$ on the spacetime ($\mathcal{M},\mathbf{g}$)
satisfying 
\begin{equation}
\mathcal{L}_{\mathbf{H}}g_{AB}=2\psi g_{AB}  \label{SelfCondition1}
\end{equation}%
where $\psi =$const. is the \emph{homothetic factor} and essentially
represents the \emph{constant} scale amplification or dilation of the
geometrical and dynamical variables. The properties underlying a
self-similar $\left( d+1\right) -$dimensional spacetime depend on whether
the homothetic vector field possesses a zero or a fixed point $p$ such that $%
\mathbf{H}(p)=0$ and $p\in \mathcal{M}$. In particular, for $d=4$ it is a
well known result that the existence of a zero $\mathbf{H}(p)=0$ implies
that the spacetime is of Petrov type III or N or O and the Ricci tensor has
a Segr\'{e} type [$(2,11)$] or [$(3,1)$] at $p$ \cite{HallBook1}. In
addition, it can be proved straightforwadly that when $\mathbf{H}(p)=0$ for
some $p\in \mathcal{M}$ then the homothetic bivector $F_{ab}(\mathbf{H}%
)\equiv \frac{1}{2}\left( H_{a;b}-H_{b;a}\right) $ is well-defined and \emph{%
nowhere zero} in ($\mathcal{M},\mathbf{g}$) (if it was zero then $\mathbf{H}=%
\mathbf{0}$ everywhere in $\mathcal{M}$). These features also describe
the general case of a $\left( d+1\right) -$dimensional spacetime by using
exactly the same method as in the 4D models. In some sense, the above
consequences of the existence of a zero are undesirable since in most cases
they impose severe restrictions on the geometry and the dynamics of the
associated model leading either to physically unsound models or models with
very special (and hence unrealistic) properties.

On the other hand, by relaxing the existence of a zero for the homothetic
vector field i.e.  $\mathbf{H}\neq \mathbf{0}$ everywhere in $\mathcal{M}$
the situation is changed dramatically because in this case we can always
choose a coordinate system adapted to $\mathbf{H}$ which is usually referred
as \emph{geodesic} coordinates. The metric takes the form 
\begin{equation}
ds^{2}=g_{AB}dx^{A}dx^{B}=e^{2\psi w}\hat{g}_{AB}(x^{\gamma })dx^{A}dx^{B}
\label{Geodesic1}
\end{equation}%
where $\mathbf{H}=\partial _{w}$ and the $\hat{g}_{AB}(x^{\gamma })$ does
not depend on $w$ hence it admits $\mathbf{H}$ as Killing Vector Field
(KVF). In case where the bulk spacetime is Locally Rotationally Symmetric
(LRS) \cite{Apostol-Carot} then it is straightforward to generalize the
4-dimensional result stating that the quantity $w$ represents a
dimensionless scale invariant variable \cite{Carr-Coley}. Although in models
where the rotational symmetry is broken this property of the coordinate $w$
is far from obvious, however we shall continue to assume that $w=t/r$
essentially corresponds to the ratio of some space $r$ and time $t$ scales
hence the dimensionless quantities of the bulk spacetime are scale invariant.

We note that in the literature the self-similar metric (\ref{Geodesic1}) is
referred as the \emph{Einstein frame} and the $\hat{g}_{AB}(x^{\gamma })$ as
the \emph{physical }or\emph{\ Jordan frame}. Still, the resulting dynamical
equations maintain their formidable complexity and one should make some
additional geometric assumptions e.g. the existence of extra symmetries for $%
g_{AB}$ or that the spacetime is algebraically special, losing in that way
features of the general case. Nevertheless the nowhere vanishing of $\mathbf{%
H}$ permit us to make a different type of simplification of the general
problem. In fact, if the Petrov type of the Weyl tensor is constant and the
Ricci tensor has the same Segr\'{e} type in all over the spacetime it
follows that we can assume without any loss of generality the vanishing of
the homothetic bivector $F_{ab}(\mathbf{H})=0$ on $\mathcal{M}$ \cite%
{HallBook1}. In this case the most general form of the $\left( d+1\right) -$%
dimensional metric is 
\begin{equation}
ds^{2}=e^{2\psi w}\left[ dw^{2}+\tilde{g}_{\alpha \beta }(x^{\gamma
})dx^{\alpha }dx^{\beta }\right]  \label{NormalGeodesic1}
\end{equation}%
i.e. $g_{AB}$ is conformally related to a $\left( d+1\right) $ \emph{global
decomposable} spacetime $\hat{g}_{AB}$ admitting the \emph{covariantly
constant} vector field $\mathbf{\hat{H}}=\partial _{w}$ and $\tilde{g}%
_{\alpha \beta }(x^{\gamma })$ is the metric of the $d-$dimensional geometry.

Then, from the Ricci identities we get 
\begin{equation}
R_{AB}H^{B}=0=\hat{R}_{AB}H^{B}  \label{Ricci1}
\end{equation}%
where $R_{AB},\hat{R}_{AB}$ are the Ricci tensors of the Einstein and
physical frame respectively.

Furthermore, equation (\ref{NormalGeodesic1}) implies that $R_{AB},\hat{R}%
_{AB}$ are related via the remarkably simple relation 
\begin{equation}
\hat{R}_{AB}=R_{AB}+\left( d-2\right) \psi ^{2}\left( \hat{g}_{AB}-\hat{H}%
_{A}\hat{H}_{B}\right) .  \label{RicciRelation1}
\end{equation}%
Using (\ref{Ricci1}), equation (\ref{RicciRelation1}) gives 
\begin{equation}
\tilde{R}_{\alpha \beta }=R_{\alpha \beta }+\left( d-2\right) \psi ^{2}%
\tilde{g}_{\alpha \beta }.  \label{RicciRelation2}
\end{equation}%
Consequently, assuming a specific matter configuration for the self-similar $%
\left( d+1\right) -$dimensional bulk spacetime in the Einstein frame, one
ends up with an \emph{effective }$d-$\emph{dimensional theory with a
positive cosmological constant which is determined by the homothetic factor }%
$\psi $ in the Einstein frame.

Clearly the above construction is closely related with the recently proposed
method in ref. \cite{Townsend} (see also \cite{PoncedeLeon:2008ai} for a
treatment using a similar mechanism in a spherically symmetric geometry and 
\cite{PoncedeLeon:2007jj}, \cite{Rippl:1995ii}) where an appropriate
compactification of the $n$ internal dimensions leads to an effective 4D
theory in the physical frame which can have both accelerating and
decelerating periods. However there are two important differences:\ first we
have given a clear geometrical interpretation of the $\left( d+1\right) $
Einstein frame as representing a general self-similar metric. On the other
hand, it is well known that self-similar models have sound physical
significance since they describe (very often) the asymptotic states of more
general models. This fact has lead to the "similarity hypothesis" stating
that under certain physical circumstances which have never been precisely
specified, general cosmological or astrophysical solutions will naturally
evolve to a self-similar model \cite{Carr:2005uf}. Therefore we expect that
self-similar $\left( d+1\right) -$dimensional bulk solutions will play a
similar role like the 4d counteparts hence this expectation is in full
agreement with the fact that scale-invariant models arise as \emph{%
equilibrium} or \emph{fixed} points in the dynamical state space and may act
as past or future attractors of general models \cite{Wainwright-Ellis}, \cite%
{Coley-Book}. In addition, in the majority of the problems of physical
interest, the HVF is always a gradient vector field (see e.g. \cite{Apostol1}%
-\cite{Apostol4} for a 4D treatment) hence $F_{AB}(\mathbf{H})=0$ (for the
case of a (proper) Conformal Vector Field, it has been shown recently \cite%
{Apostolopoulos:2022mcz} the significance of the vanishing of the conformal
bivector in constructing viable scalar field models in a spherically
symmetric 4D spacetime).

In this sense and assuming that these heuristic arguments are true for
general bulk matter configurations, the metric (\ref{NormalGeodesic1}) maybe
represents the \emph{most general solution of a }$\left( d+1\right) -$\emph{%
dimensional model at early or late times}. In addition, we have identified
the origin of the positive cosmological constant in the $d-$dimensional
spacetime as the homothetic factor hence \emph{a geometrical artifact of
late time state} of the $\left( d+1\right) -$dimensional bulk spacetime.

A second difference with \cite{Townsend} is that we do not assume a Ricci
flat $\left( d+1\right) -$dimensional bulk spacetime since, in the spirit of
the above mentioned late time behaviour and using equation (\ref%
{RicciRelation2}), this will always lead to an empty effective $d-$%
dimensional model with a positive cosmological constant. In fact this is
exactly the situation with the Kasner-type models considered in \cite%
{Townsend} which always admit a proper HVF \cite{Tsamp-Apostol4}
irrespective the dimension. In contrast, we have a non-empty bulk $%
R_{AB}\neq 0$ giving rise to a \emph{non-empty} $d-$dimensional spacetime.
Consider for example a $\left( d+1\right) -$dimensional bulk energy-momentum
tensor of the form 
\begin{equation}
T_{AB}=\left( \rho +p\right) u_{A}u_{B}+p\left( g_{AB}-e_{A}e_{B}\right)
\label{EnergyMomentum1}
\end{equation}%
with $e^{A}=e^{-\psi w}\delta _{w}^{A}$ a global nowhere zero smooth vector
field and $u_{A}e^{A}=0$. The above type of bulk energy-momentum can be
interpreted as an anisotropic fluid in the $\left( d+1\right) -$dimensional
spacetime or equivalently as \textquotedblleft perfect\textquotedblright\
fluid confined in a $d-$dimensional hypersurface. To see this more clear we
remark that the existence of the homothety implies that $\rho =\hat{\rho}%
(x^{\alpha })e^{-2\psi w}$ and $p=\hat{p}(x^{\alpha })e^{-2\psi w}$ for some
smooth \emph{positive} functions $\hat{\rho}(x^{\alpha }),\hat{p}(x^{\alpha
})$. Hence in the Einstein frame the bulk energy-momentum (\ref%
{EnergyMomentum1}) is written 
\begin{equation}
T_{AB}=\left[ \hat{\rho}(x^{\alpha })+\hat{p}(x^{\alpha })\right] \hat{u}_{A}%
\hat{u}_{B}+\hat{p}(x^{\alpha })\left( \hat{g}_{AB}-\delta _{A}^{w}\delta
_{B}^{w}\right) .  \label{EnergyMomentum2}
\end{equation}%
Using (\ref{Ricci1}) we get $\hat{p}(x^{\alpha })=\hat{\rho}(x^{\alpha
})/(d-1)$. Then, by means of the usual Kaluza-Klein compactification scheme
the effective $n-$dimensional Einstein Field Equations in the physical frame
are 
\begin{equation}
\hat{R}_{\alpha \beta }=\hat{\rho}(x^{\alpha })\hat{u}_{\alpha }\hat{u}%
_{\beta }+\frac{1}{d-1}\hat{\rho}(x^{\alpha })\left( \hat{g}_{\alpha \beta }+%
\hat{u}_{\alpha }\hat{u}_{\beta }\right) +\left( d-2\right) \psi ^{2}\hat{g}%
_{\alpha \beta }  \label{EffectiveEFE1}
\end{equation}%
i.e. we get an \emph{effective }$d-$\emph{dimensional perfect fluid
(radiation) model plus an effective positive cosmological constant }$\Lambda
=\left( d-2\right) \psi ^{2}$.

As another example we consider the case of bulk perfect fluid 
\begin{equation}
T_{AB}=\left( \rho +p\right) u_{A}u_{B}+pg_{AB}.  \label{PerfectFluid1}
\end{equation}
Following a similar approach we obtain 
\begin{equation}
\hat{R}_{\alpha \beta }=2\hat{\rho}(x^{\alpha })\hat{u}_{\alpha }\hat{u}%
_{\beta }+\left( d-2\right) \psi ^{2}\hat{g}_{\alpha \beta }
\label{EffectiveEFE2}
\end{equation}
i.e. a \emph{stiff perfect fluid}.

Let us remark that in \emph{all} the examples considered above, there exist 
\emph{exact} and \emph{analytical} self-similar models \cite%
{Apostolopoulos5DInhomogeneous} for which the Einstein Field Equations are satisfied for the
specific bulk matter setups (\ref{EnergyMomentum1}) and (\ref%
{EnergyMomentum2}). Of course, in order to have a complete picture of the
procedure followed here, one should seek for appropriate bulk solutions and
show that a self-similar model is the future attractor. If this is the case,
then the proposal of the present article might provide an alternative
explanation of the appearance and dominance of a positive cosmological
constant at late times in the $d-$dimensional spacetime and the matter
dominated epoch at early times.\newline
\newline
\textbf{Data availability} Data sharing not applicable to this article as no
datasets were generated or analysed during the current study. \newline
\newline

\end{document}